\documentclass[preprint,prX]{revtex4}
\usepackage{hyperref}
\usepackage{graphicx} 
\usepackage{epsfig}

\begin{document}

\title{The Boson peak in supercooled water}

\author{Pradeep Kumar}
\affiliation{Center for Studies in Physics and Biology, The Rockefeller University, 1230 York Avenue, New York, NY 10021 USA }
\author{K. Thor Wikfeldt}
\affiliation{Science Institute and Faculty of Science, VR-III, 
  University of Iceland, 107 Reykjav\'ik, ICELAND}
\affiliation{Nordita, Royal Institute of Technology and Stockholm University, 
Roslagstullsbacken 23, S-10691 Stockholm, SWEDEN}

\author{Daniel~Schlesinger}
\author{Lars~G.~M.~Pettersson}
\affiliation{Department of Physics, AlbaNova, Stockholm University, 
S-10691 Stockholm, SWEDEN}
\author{H.~Eugene~Stanley}
\affiliation{Center for Polymer Studies and Department of Physics, Boston 
  University, Boston, MA 02215 USA}

\begin{abstract} 
 
We perform extensive molecular dynamics simulations of the TIP4P/2005
model of water to investigate the origin of the Boson peak reported in
experiments on supercooled water in nanoconfined pores, and in
hydration water around proteins. We find that the onset of the Boson
peak in supercooled bulk water coincides with the crossover to a
predominantly low-density-like liquid below the Widom line $T_W$. The frequency and onset temperature of the Boson peak in our
simulations of bulk water agree well with the results from experiments
on nanoconfined water. Our results suggest that the Boson peak in
water is not an exclusive effect of confinement. We further find
that, similar to other glass-forming liquids, the vibrational modes
corresponding to the Boson peak are spatially extended and are related
to transverse phonons found in the parent crystal, here ice Ih.
 
\end{abstract} 


\maketitle

\section*{Introduction:~~~} 
One of the characteristic features of many glasses and amorphous
materials is the onset\cite{Angell2004} of low-frequency collective
modes (Boson peak) in the energy range $2-10$~meV at low $T$, where
the vibrational density of states (VDOS) $g(\omega)$ shows an excess
over $g(\omega)\propto\omega^2$ predicted by the Debye
model. Disordered materials are further known to exhibit many
anomalous behaviors compared to their crystalline counterparts, such
as the temperature dependence of thermal
conductivity\cite{Cahil1987,Kumar2011} and specific
heat\cite{Berman1949,Zeller1971} at low temperatures. Many
scenarios\cite{nakayama2002Boson,angell2000relaxation} have been
suggested to explain the physical mechanisms behind the Boson peak and
related anomalies, but a comprehensive understanding has proved
elusive.

Recent neutron scattering experiments on water confined in nanopores
indicate the presence of a Boson peak\cite{chen2008dynamic,chen2009evidence}
around $5-6$~meV ($40-49$~cm$^{-1}$) emerging below 230\,K in the
incoherent dynamic structure factor. These
results were tentatively interpreted as arising from a gradual
change in the local structure of confined liquid water when crossing the Widom 
line temperature $T_W$\cite{Kumar2006}.
Earlier, neutron scattering has also been applied to protein hydration 
water~\cite{paciaroni1999} and a Boson peak was found around 30~cm$^{-1}$.
$T_W$ corresponds to the loci of maxima of thermodynamic response 
functions in the one-phase region beyond the liquid-liquid critical point 
(LLCP) proposed to exist in supercooled liquid water\cite{poole1992phase}.  
A Widom line in the supercritical region in argon has recently been
studied\cite{Simeoni2010,gorelli2013} and found to be directly related
to a dynamical crossover between liquid-like and gas-like properties,
but the existence of a dynamical crossover in supercooled water is
subject to some
controversy~\cite{faraone2004fragile,mamontov2005observation,swenson2006relaxation,mallamace2006fragile,swenson2007properties,khodadadi2008origin,pawlus2008conductivity,vogel2008origins,vogel2009temperature}.
Since the melting temperature is strongly depressed in nanoconfinement and
in protein hydration water, deeply supercooled confined water has been 
used experimentally to infer the behavior of 
bulk water, which is more challenging to supercool. 
However, the similarity to bulk water has been called into 
question\cite{ricci2000,mattea2008dynamics}, and it is thus important to 
further investigate the relationship between water in these different forms.

Both experimental\cite{chen2006observation,chen2009evidence} and
simulation\cite{xuPNAS} studies suggest that the density relaxation
of confined and hydration water at and slightly below the Widom
temperature is of the order of a few tens of nanoseconds, implying
that liquid water is still in metastable equilibrium over the
experimental time scales involved\cite{Liu2005}.  The experimental
observation of a Boson peak below $T_W$ in confined water thus
suggests that the low-density-like liquid shares vibrational
properties with the glassy state, as has been observed previously for
other systems such as $B_2O_3$\cite{borjesson1993,sokolov1994}. On the
other hand, the dynamic structure factor of crystalline ice exhibits a
peak at a slightly higher frequency around
$50$~cm$^{-1}$\cite{bermejo1995absence,yamamuro2001low,yamamuro2003inelastic}
as do the Raman spectra of ice Ih and proton-ordered ice
XI\cite{Kohji2011}; in the latter case the peak becomes extremely
sharp. These results indicate a connection between the dynamics of
supercooled liquid and crystalline vibrational dynamics. Indeed, it
has recently been suggested that Boson peaks observed in glasses are
related to the vibrational dynamics of the parent
crystal\cite{Chumakov2011}.

To shed light on the question of whether bulk supercooled liquid water
displays a Boson peak, we study the low-frequency dynamics of the
TIP4P/2005 model of water which accurately reproduces a range of water
properties~\cite{abascal2005tip4p2005}, including its
anomalies~\cite{pi2009anomalies}. This model has been found to exhibit
a LLCP in the vicinity of $P_C=1350$~\,bar and
$T_C=193$~\,K\cite{abascal2010widom}. The associated Widom line has
been shown to be accompanied by a structural crossover from a
predominantly high-density liquid (HDL) structure to a predominantly
low-density liquid (LDL) structure\cite{WikfeldtJCP,WikfeldtPCCP}
occurring at a temperature $T_W$ around $230$~\,K at 1~\,bar. We find
that this water model indeed displays a Boson peak in the bulk
supercooled regime and that its onset coincides with the Widom
temperature. Our analysis further shows that it derives from
transverse acoustic modes in the parent crystal, ice Ih, in agreement
with the recently proposed picture\cite{Chumakov2011}. We further verify our results 
using another model of water, TIP5P, and find also in this case an emergence of a Boson peak below the Widom line.

\section*{Results}
\subsection{Incoherent dynamic structure factors:~~~}

A quantity that is readily accessible in inelastic neutron scattering
experiments is the incoherent dynamic structure factor (DSF)
$S_S(k,\omega)$ which probes single-particle dynamics. To compare our
simulation results with experiment we first calculate the
self-intermediate scattering function
\begin{equation} F_S(k,t) = \left< 
\frac{1}{N}\sum_{i}e^{i{\bf k}\cdot[{\bf r}_i(t)-{\bf 
      r}_i(0)]}\right>  
\label{eq:fkt} 
\end{equation} 
where $\mathbf{r}_i(t)$ are the positions of oxygen atoms, $N$ is the number 
of molecules and angular brackets denote an ensemble average and averaging over
different ${\bf k}$ with the same modulus. 
In simulations the wave vector $\mathbf{k}$ is defined as
$\frac{2\pi}{L}(n_x,n_y,n_z)$ for integers $(n_x,n_y,n_z)$ and system 
size $L$.
We perform the frequency
decomposition of $F_S(k,t)$ to obtain the incoherent DSF
\begin{equation} 
S_S(k,\omega) \equiv \int F_S(k,t)e^{i\omega t}dt. 
\end{equation} 

\begin{figure*}[htb] 
\begin{center} 
\includegraphics[width=16cm]{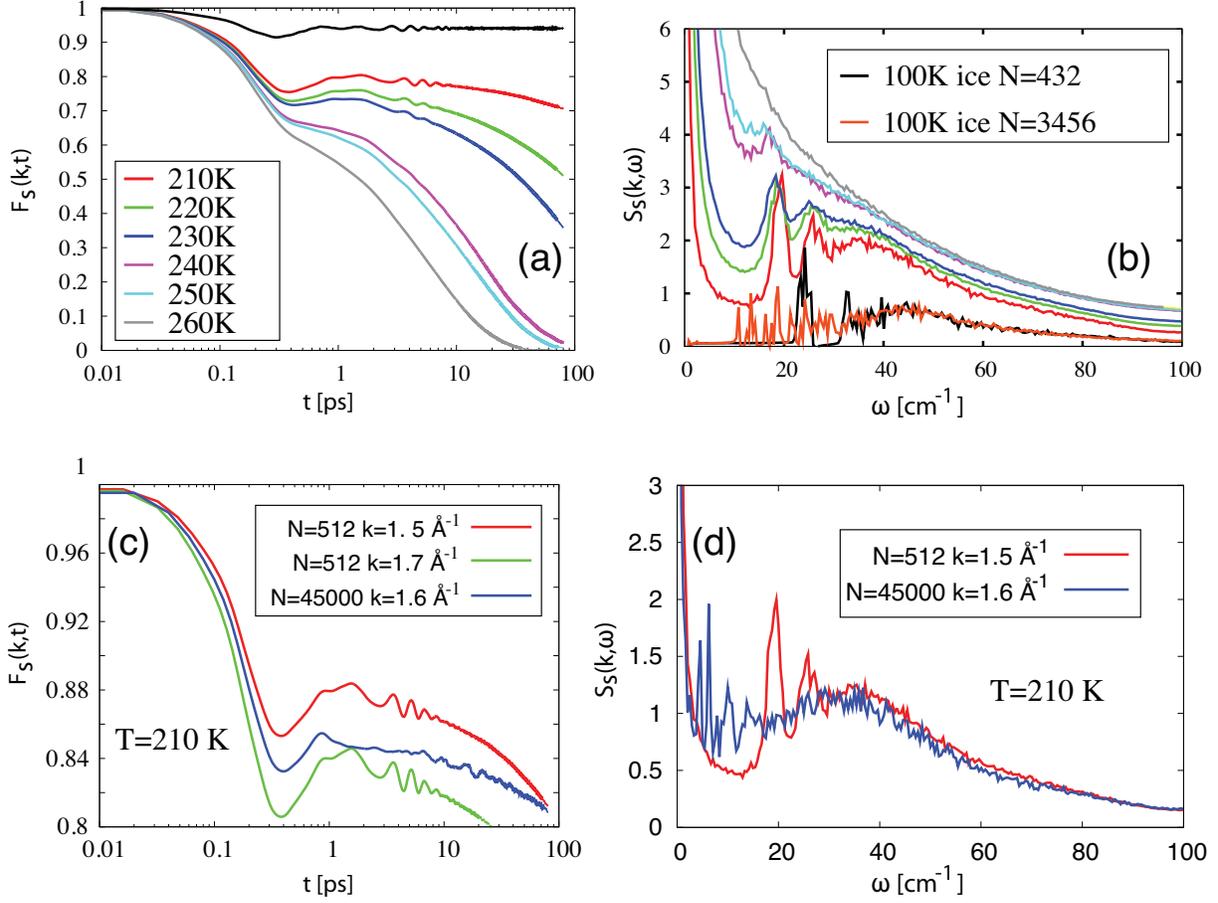} 
\end{center} 
\caption{(a) $F_s(k,t)$ and (b) $S_s(k,\omega)$ for a
  range of temperatures 210-260~K at fixed $k=2.0$~{\AA}$^{-1}$ and system
  size $N=512$.  $S_S(k,\omega)$ at $k=2.0$~{\AA}$^{-1}$ for ice at
  100~K is also shown for two different system sizes, $N=432$ and
  $N=3456$.  (c)-(d) show a comparison with $F_s(k,t)$ and
  $S_S(k,\omega)$ for a large $N=45,000$ simulation at 210~K at
  k$\approx$1.6~{\AA}$^{-1}$. The broad peak around
  $\omega=37$~cm$^{-1}$ is evidently independent of system size while
  the sharper peaks at lower frequency are not present in the large
  simulation.}
\label{fig:fig1} 
\end{figure*}

In Figs.~\ref{fig:fig1}(a) and \ref{fig:fig1}(b), we show $F_S(k,t)$ and
$S_S(k,\omega)$ for supercooled liquid water simulations with $N=512$
molecules at atmospheric pressure and different temperatures for
$k=2$~\AA$^{-1}$, {\it i.e.}, the position of the first peak in the 
structure
factor $S(k)$.  For $T<T_W$, a minimum appears in $F_S(k,t)$ around
$0.3$~ps followed by oscillations up to around $10$~ps. In the incoherent 
DSF these changes with temperature correspond
to the emergence of sharp peaks at $20$ and $25$~cm$^{-1}$ along with a
broad peak centered around $37$~cm$^{-1}$. For comparison, we also show
$F_S(k,t)$ and $S_S(k,\omega)$ from hexagonal ice simulations at 100~K
for two different system sizes. The connection between liquid and
crystalline low-frequency dynamics will be further discussed below but
we note here that the system size obviously affects the region
$\omega<40$~cm$^{-1}$.

Since low-frequency dynamics in supercooled liquids and glasses may be
affected by finite-size
effects~\cite{horbach1996finite,kim2000apparent,horbach2001high}, we
show in Figs.~\ref{fig:fig1}(c) and \ref{fig:fig1}(d) a comparison
with a much larger system, $N=45,000$ at 210~K. In $F_S(k,t)$ the
oscillations are shifted to longer times while the minimum at
$0.3-0.4$~ps and subsequent peak around $0.9$~ps are system-size
independent. The sharp peaks in $S_S(k,\omega)$ vanish in the large
simulation but the broad peak around $37$~cm$^{-1}$ persists; we
therefore label this peak of low-frequency excitations as the Boson
peak.  Comparing our simulation results with experimental neutron data
on protein hydration water\cite{paciaroni1999} and on water in
confinement\cite{chen2008dynamic,chen2009evidence} reveals rather good
agreement. Both the experimental energy position, ~$30$~cm$^{-1}$ for
hydration water\cite{paciaroni1999} and $45$~cm$^{-1}$ for confined
water\cite{chen2008dynamic,chen2009evidence}, and the temperature
onset, $T=225$~K\cite{chen2008dynamic,chen2009evidence}, are
surprisingly well reproduced by the TIP4P/2005 model considering the
approximate nature of classical force fields. Since we observe
well-defined low-frequency modes in the incoherent DSF in the
simulations of bulk water, we conclude that the Boson peak in
supercooled water is not a consequence of confinement and that it
would likely be detected also in experiments on bulk water, if
sufficient supercooling could be achieved.

A connection to the Widom line in TIP4P/2005 is clearly present since
a qualitative change of both $F_S(k,t)$ and $S_S(k,\omega)$ takes
place between 230 and 240~K. Our results thus imply that a Boson peak
may also appear in other tetrahedral~\cite{Errington2001} liquids, such as silicon and
silica, for which simulations have indicated the existence of a
liquid-liquid phase transition\cite{sastry2003,Saika2000}.  To confirm the
connection between the Boson peak and the Widom line we have performed
additional simulations of a different water potential, the TIP5P
model\cite{mahoney2000five}, which has been shown to exhibit a Widom
line around 250~K\cite{xuPNAS}, 20~K above that for TIP4P/2005. As
shown in the Supplementary Information, a Boson peak emerges also in
this model and, indeed, it coincides with $T_W$ at 250~K.

\subsection{Vibrational density of states:~~~} 

The Boson peak is commonly discussed in terms of the VDOS,
$g(\omega)$. A distinct peak is often not seen in the VDOS itself but
appears only in the reduced VDOS after normalizing by the squared
frequency, $g(\omega)/\omega^2$, which reveals the excess over the Debye
model prediction $g(\omega)\propto \omega^2$. We calculate $g(\omega)$
as the Fourier transform of the oxygen velocity autocorrelation
function, $C_v(t)$:
\begin{equation} 
g(\omega) \equiv \int C_v(t)e^{i\omega t}dt, 
\label{eq:vdos}
\end{equation} 
where 
\begin{equation} 
C_v(t) = \left< \frac{1}{N}\sum_{i}{\vec v}_i(t){\vec 
v}_i(0)\right>.
\end{equation}  
The sum includes all $N$ oxygens in the system, ${\vec v}_i$ are the
oxygen atom velocities and
$\langle\ldots\rangle$ denotes the ensemble average.
  
\begin{figure*} 
\begin{center} 
\includegraphics[width=16cm]{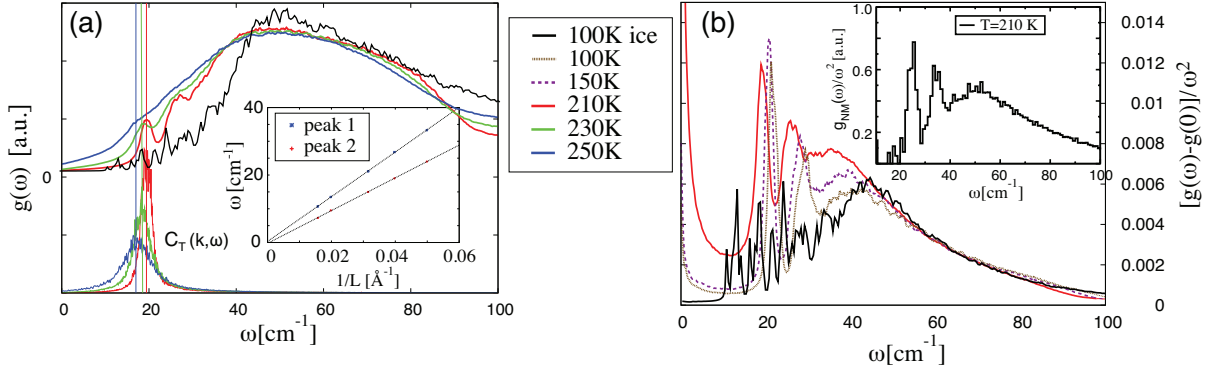} 
\end{center} 
\caption{Simulations of TIP4P/2005 water with
  $N=512$. (a) VDOS of liquid water at supercooled temperatures and of
  hexagonal ice and (inset) inverse box-length dependence of the two
  lowest sharp peaks of the VDOS, extrapolating to zero frequency in the
  limit of infinite system size. In (a) the lowest-$k$ transverse
  current spectrum is shown in the bottom part, illustrating that the
  sharp low-frequency peaks are low-$k$ transverse modes. (b) Reduced
  vibrational density of states at low $T$ calculated as
  $\left[g(\omega)-g(0)\right]/\omega^2$ with an extrapolated $g(0)$ to 
  eliminate uncertainties related to the finite simulation time. 
  Upon rapid cooling into a non-equilibrated LDA ice at 150~K and 100~K, 
  the Boson peak is seen to
  shift to higher frequencies, approaching that of hexagonal ice. The
  inset in (b) shows the reduced VDOS calculated from the normal modes
  of inherent structures quenched from equilibrated T=$210$~K
  configurations.  The normal mode $g_{NM}(\omega)/\omega^2$ shows a
  Boson peak which is blue-shifted to about the same frequency as the
  crystal, suggesting that as the liquid structure is made harmonic the
  Boson peak frequency shifts to higher values saturating at about
  $50$~cm$^{-1}$.}
\label{fig:fig2} 
\end{figure*}

In Fig.~\ref{fig:fig2}(a) we show graphs of $g(\omega)$ for various
temperatures. Below $T_W$, $g(\omega)$ shows an onset of the same two
sharp low-frequency peaks as observed in $S_S(k,\omega)$ in
Fig.~\ref{fig:fig1}.  By simulating a range of different system sizes
we can establish the system-size dependence of $g(\omega)$.  The
agreement for frequencies $\omega > 40$~cm$^{-1}$ is very good for
different system sizes L, 
but the sharp low-frequency peaks
shift to lower frequencies as the system size is increased. Indeed,
both peaks extrapolate linearly to zero as $1/L$, as
seen in the inset of Fig.~\ref{fig:fig2}(a), suggesting that they
disappear in the limit of infinite system size.

As discussed in relation to Fig.~\ref{fig:fig1}, there is a
low-frequency peak in $S_S(k,\omega)$ for hexagonal ice at somewhat
higher frequency compared to the supercooled liquid simulations,
suggesting a link between the liquid and crystalline low-frequency
modes.  We investigate this by performing thermally non-equilibrated
simulations at even lower temperatures where the TIP4P/2005 model
vitrifies to a low-density amorphous (LDA)-like solid.  
The non-equilibrium simulations are
performed by annealing the equilibrated metastable 210~K simulation with
a cooling rate of 2$\cdot$10$^{10}$~K/s to reach target temperatures of
150 and 100~K.  Figure~\ref{fig:fig2}(b) shows the reduced VDOS,
$[g(\omega)-g(0)]/\omega^2$ for simulations at 100, 150 and 210~K
compared to the hexagonal ice simulation at 100~K.  We subtract an
extrapolated value of $g(0)$ from $g(\omega)$ to eliminate uncertainties
due to the finite length of the trajectories when evaluating the Fourier
transform in Eq.~(\ref{eq:vdos}).  A clear shift to higher frequencies
of the Boson peak is observed as the supercooled liquid simulations are
cooled into the LDA glass region, and the reduced VDOS at 100~K
resembles the crystalline ice counterpart.

In the inset of Fig.~\ref{fig:fig2}(b), we show the reduced VDOS
$g_{NM}(\omega)/\omega^2$ obtained from the normal modes of the
liquid calculated from quenched configurations (\textit{i.e.}, energy 
minimized inherent structures) at
$T=210$~K.  We find that the Boson peak in $g_{NM}(\omega)/\omega^2$ is
blue-shifted to around $50$~cm$^{-1}$, close to the peak for hexagonal ice, 
compared to the velocity autocorrelation function VDOS which peaks 
at $37$~cm$^{-1}$. This suggests that a key difference in low-frequency 
vibrational properties between the LDL-like liquid below the Widom line 
and crystalline ice lies in the more anharmonic dynamics of the liquid phase.

\subsection{Transverse and longitudinal correlation functions:~~~}

Having established the existence of a Boson peak in supercooled water
below the Widom line and its connection to low-frequency dynamics
present in the parent crystal, ice Ih, we now turn to the study of
transverse and longitudinal current correlations to clarify the nature
of these low-frequency modes.  We calculate longitudinal and
transverse currents as
\begin{eqnarray} 
J_L(\mathbf k,t) &=& \sum_{i=1}^N\mathbf{\hat k_{\parallel}\cdot  
v}_i(t)\exp 
\left[ -i \mathbf k \cdot \mathbf r_i(t) \right]\, , \\ 
J_T(\mathbf k,t) &=& \sum_{i=1}^N\mathbf{\hat k_{\perp}\cdot 
  v}_i(t)\exp\left[ -i \mathbf k \cdot \mathbf r_i(t) \right]  
\end{eqnarray} 
where $\mathbf {\hat k_{\parallel}}$ and $\mathbf {\hat k_{\perp}}$ are 
unit vectors respectively parallel and perpendicular to $\mathbf k$, and 
$\mathbf{r}_i(t)$ and $\mathbf{v}_i(t)$ denote the oxygen atoms' position and 
velocity, respectively. The frequency 
decomposition of the longitudinal and transverse current autocorrelation 
functions is 
\begin{equation} 
C_{\alpha}(k,\omega) \equiv
\int \langle{J}_{\alpha}(k,t){J}_{\alpha}(-k,0)\rangle 
e^{i\omega t} dt 
\label{eq:currentspectra} 
\end{equation} 
where $\alpha=L$~or $T$. 

In the bottom part of Fig.~\ref{fig:fig2}(a) we show superimposed on the
VDOS the transverse current correlation function (TCCF) $C_T(k,\omega)$ 
at different temperatures for the lowest
wave number $k$ accessible in the $N$=$512$ simulation boxes, {\it i.e.},
$\mathbf{k}=2\pi\left( 1,0,0\right)/L$ and permutations thereof.  We see
that the first sharp size-dependent low-frequency peak in VDOS, which develops
below $T_W$, coincides exactly with the lowest-$k$ TCCF, and the second
finite-size peak around $25$~cm$^{-1}$ coincides with the
second-lowest-$k$ TCCF (not shown). Returning to Fig.~\ref{fig:fig1}(d), the
low-frequency side of the Boson peak is smoother in the large $N=45,000$
simulations and the sharp peaks related to the lowest-$k$ transverse
currents are instead seen at frequencies below 10~cm$^{-1}$. We thus
conclude that the sharp system size dependent low-frequency peaks around
the Boson peak frequency and below are predominantly transverse
excitations, consistent with previous findings for amorphous 
silica~\cite{horbach2001high}.

The dispersion relations for transverse and longitudinal current
spectra, $C_T(k,\omega)$ and $C_L(k,\omega)$ at $T=210$~K, and
$P=1$~atm are shown in Fig.~\ref{fig:fig4} (the current spectra are
shown in the Supplementary Material). We obtain dispersion relations by fitting damped harmonic oscillator (DHO)~\cite{Sette2008} lines to both longitudinal and transverse spectra.

At small $k$ below
0.5~\AA$^{-1}$, the longitudinal current spectrum, $C_L(k,\omega)$,
shows only one acoustic dispersing excitation. For $k>0.5$~\AA$^{-1}$,
$C_L(k,\omega)$ shows the existence of three excitations and can be fit with three DHO lines. Besides the
dispersing excitation at intermediate frequency, two other
nondispersing excitations appear---one at low frequency around
50~cm$^{-1}$ and the other at high frequency around 260~cm$^{-1}$. The
intensity of these excitations in $C_L(k,\omega)$ increases upon
further increase of $k$. Transverse current spectra $C_T(k,\omega)$
exhibit an acoustic dispersing excitation $k<0.5$~\AA$^{-1}$. For $k>0.5$~\AA$^{-1}$,
$C_T(k,\omega)$ develops a peak at $\omega \approx 260$~cm$^{-1}$, the
same excitation as in the longitudinal current spectra. Transverse and longitudinal modes are thus strongly mixed above
0.5~\AA$^{-1}$ as has been found previously also for the SPC/E water
model~\cite{sampoli1997}. Moreover, the mixing happens at both high frequencies ($\omega \approx 260$~cm$^{-1}$) and low frequencies ($\omega < 60$~cm$^{-1}$) as evident from transverse excitations appearing in longitudinal spectra. We note that the band at $260$~cm$^{-1}$ in liquid water is associated
with four-coordinated water molecules since low-frequency Raman
spectra of water down to $-20^{\circ}$~C showed its intensity to
increase with decreasing temperature\cite{Krishnamurthy1983} and
hexagonal ice also displays a strong band near this
frequency~\cite{klug1991}. Hence, the emergence of the $260$~cm$^{-1}$ band in both $C_L(k,\omega)$ and
$C_T(k,\omega)$ suggests that liquid water at this temperature
exhibits networks of four-coordinated molecules over length scales as
large as $2\pi/0.5$~\AA$^{-1}\sim 13$~\AA. 
This observation is also
consistent with recent studies where it is shown that the sizes of
clusters of highly tetrahedral molecules increases below the Widom
temperature\cite{Kumar2011,Errington2001}.

The emergence of an additional, high-frequency excitation in
$C_L(k,\omega)$ and $C_T(k,\omega)$ around $260$~cm$^{-1}$ for
$k>0.5$~\AA$^{-1}$ suggests a low-frequency liquid-like and a
high-frequency solid-like response of the longitudinal and transverse
spectra at this length scale and a concomitant pile-up of spectral intensity takes place in the Boson peak regime.

To get further insights into the longitudinal or transverse character
of the Boson peak, we compare cumulatively integrated spectra over $k$ 
for different frequencies for both the transverse and longitudinal parts, $C^{*}_{L}(k,\omega)$
and $C^{*}_{T}(k,\omega)$,
defined as
\begin{equation}
\label{eq:intcurr}
C^{*}_{\alpha}(k,\omega) = \int_{k_{\rm min}}^{k}C_{\alpha}(k',\omega)dk', 
\end{equation}
where $\alpha=L$~or $T$ and $k_{\rm min}=2\pi/L$ is the smallest
wave number accessible in our system of box size
$L$. $C^{*}_{\alpha}(k,\omega)$ thus describes the total contribution of
longitudinal and transverse modes with different wave numbers up to $k$
for a given frequency $\omega$. In Fig.~\ref{fig:fig5}(a)-(b) we show
$C^{*}_{L}(k,\omega_B)$ and $C^{*}_{T}(k,\omega_B)$ for several
frequencies $\omega_B$ in the Boson peak region around 37~cm$^{-1}$.
It can be clearly seen that transverse modes are dominant in the Boson
peak frequency regime for $k>0.5$~\AA$^{-1}$.

\subsection{Localization analysis:~~~} 
 
A number of studies on amorphous materials have found that the modes
in the Boson peak frequency range are localized or
quasi-localized\cite{McIntosh1997,Novikov1995}. In order to investigate
whether this holds also for TIP4P/2005 water below the Widom line
$T_W$ we calculate the degree of localization by performing a normal
mode analysis of quenched (or inherent) structures, obtained by
energy-minimizing snapshots from simulation trajectories.  A measure of
the degree of localization of a vibrational mode is the
frequency-dependent participation ratio\cite{Bell1970,Sastry1994}
$p_\mu$ for mode $\mu$
\begin{equation}  
p_\mu = \left[N\sum_{i=1}^{N}(u_\mu^i.u_\mu^i)^{2}\right]^{-1}  
\end{equation}  
where $u_\mu^i$ is the contribution of all degrees of freedom of
molecule $i$ to the normal mode $\mu$. The participation ratio is
unity when all molecules contribute equally to the normal mode in
consideration, while $p=1/N$ if only one molecule contributes to the
total energy of the mode. Hence, for an extended mode $p_{\mu}$ is
quite large and does not depend on the system size while for a
localized mode it is small and scales with system size as $1/N$. One
way to determine the extent of localization is thus to inspect the
system-size dependence of the participation ratio. We find (see Fig.~\ref{fig:fig7}(a)) 
that the participation ratio is quite large,
around $0.6$, for the modes with frequency below $50$~cm$^{-1}$, the
region of the Boson peak.  Apart from the sharp finite-size peaks, the
participation ratio for the modes in the region of the Boson peak 
show only a weak system size dependence, suggesting that the modes 
giving rise to the Boson peak are not localized but extended. 

We next introduce a function $A_{\rm max}(r)$ defined as the maximum
displacement of molecules at a distance $r$ from the molecule with the
largest displacement in the normal mode.  For a localized mode a rapid
decay of $A_{\rm max}(r)$ should be seen. In Fig.~\ref{fig:fig7}(b) we show
average $A_{\rm max}(r)$ for the normal modes in two different frequency
regimes -- for the Boson peak regime ($\omega_B=45\pm2.5$~cm$^{-1}$,
note that the frequency of the Boson peak for the quenched configuration
is shifted to higher frequency compared to the Boson peak frequency
$\omega_B\approx 37$~cm$^{-1}$ of the liquid at $T=210$~K, see inset of 
fig.~\ref{fig:fig2}(b)), and in
comparison we also show average $A_{\rm max}(r)$ for modes in the range
of $\omega = 400\pm5$~cm$^{- 1}$, which is a range of localized
vibrations. While for $\omega= 400\pm5$~cm$^{-1}$, $A_{\rm max}(r)$
decays rapidly to zero, for the Boson peak region it does
not, suggesting again an extended nature of the Boson peak modes.

\section*{Discussion~~~} 

The low-frequency vibrations of a classical potential model of water,
TIP4P/2005, are investigated in the supercooled temperature regime to
clarify the origin of the Boson peak reported from inelastic neutron
scattering experiments below around 225~K in
nano-confined\cite{chen2008dynamic,chen2009evidence} and
protein-hydration water\cite{paciaroni1999}.  We find that sharp
low-frequency peaks emerge in the incoherent dynamic structure factor and the reduced density of states of the
simulated liquid water as the system is cooled below the Widom line,
but a system-size investigation reveals that in the limit of an
infinitely large simulation box these peaks extrapolate to zero
frequency. The sharp finite-size peaks are seen to coincide exactly
with the inherently discrete, low wave-number acoustic branch of the
transverse current correlation functions, reflecting a strong
contribution of transverse modes in this frequency region.  However,
we find a broad Boson peak centered around $37$~cm$^{-1}$  which is unaffected by system size, and for which the
frequency region and temperature onset in the incoherent DSF agree
well with neutron experiments on confined water. Due to its lower
melting temperature, water in confinement has been used experimentally
to infer the behavior of bulk water below the bulk homogeneous
nucleation temperature. The validity of this comparison has been questioned, but the good agreement observed here in the
low-frequency vibrational dynamics lends support to the view of
confinement as useful in the study of supercooled bulk water, at least
for low-frequency vibrational properties.

The frequency of the Boson peak in supercooled TIP4P/2005 water as
observed in the reduced VDOS changes as the
simulation is annealed into the LDA glass region and 
approaches $\omega=45$~cm$^{-1}$. This is also the frequency at which
hexagonal ice simulations display a peak in vibrational spectra
deriving from the transverse acoustic branch, as has been observed
experimentally\cite{bermejo1995absence,yamamuro2001low,
yamamuro2003inelastic}. Thus, upon lowering the temperature below the Widom line, the
low-frequency dynamics of the system progressively changes from
LDL-like to LDA glass and to the dynamics found in hexagonal ice. A
similar shift to higher frequencies is observed in normal-mode spectra
of inherent structures quenched from liquid at temperatures below the
Widom line, indicating that the lower frequency of the Boson peak in
the liquid below the Widom line, compared to the transverse
acoustic peak in the ice, is a result of the more anharmonic nature of
the vibrational modes in the liquid.

Recent work by Chumakov {\it et al.}\cite{Chumakov2011} on glasses
suggests that there is no excess in the actual number of states at the
Boson peak and hence no additional modes compared to the crystal. The
Boson peak is thus related to the transverse acoustic singularity of
the underlying crystal structure.  Transverse modes have also been
firmly connected to the Boson peak in other
works~\cite{horbach2001high,Shintani2008}. Indeed, our studies of
transverse and longitudinal correlation functions suggest that
low-frequency transverse phonons contribute the most to the Boson peak
intensity in the range of wave numbers where both the longitudinal and
transverse phonons show a solid-like response over large length
scales, namely emergence of a high-frequency peak in both longitudinal
and transverse spectra at $\omega\approx 260$~cm$^{-1}$ for
$k>0.5$~{\AA}$^{-1}$. The appearance of this high-frequency excitation
associated with four-coordinated molecules in longitudinal and
transverse spectra coincides with a pile up of intensities in the
Boson peak regime.

To conclude, our results indicate that liquid water displays a Boson
peak below the Widom line temperature $T_W$. Both the onset
temperature and energy position are similar to what has been observed
experimentally for confined
water~\cite{chen2008dynamic,chen2009evidence}.  We find that as the
liquid crosses over to a low-density-like liquid structure below $T_W$ the
low-frequency dynamics of the liquid changes to resemble that of the
underlying crystal, ice Ih. The Boson peak in supercooled water is
thus a manifestation of the transverse acoustic singularity of the
crystal and may therefore be a general phenomenon in tetrahedral
liquids showing a liquid-liquid phase transition.

\section*{Methods} 
 
We simulate TIP4P/2005\cite{abascal2005tip4p2005} water for a range of
temperatures at atmospheric pressure. Equilibration is first performed
in the NPT ensemble, using the Nos\'{e}-Hoover thermostat and
Parrinello-Rahman barostat to attain constant temperature and
pressure. The equilibrated densities are then used in equilibration NVT
runs performed over multiple structural relaxation times, after which we
switch to the NVE ensemble to compute the relevant dynamical
quantities. The equations of motion are integrated with a time step of
0.2-1.0~fs, depending on the observed energy conservation. Most
simulations are performed using $N$=512 molecules, but to quantify the
finite-size effects we simulate larger systems up to $N$=45,000.
Simulation temperatures between 210 and 260~K at $P=1$~atm were chosen
so that the system crosses $T_W$, the temperature where maxima in
response functions are observed. 
To confirm the connection between the Widom line $T_W$ and the onset 
temperature of the Boson peak we repeat the above simulation protocol for 
the TIP5P water model\cite{mahoney2000five} at temperatures between 
240 and 270~K (see Supplementary Information).

We perform non-equilibrium simulations of TIP4P/2005 
at 100 and 150~K by rapidly cooling from 210~K with a cooling rate of
2$\cdot$10$^{10}$~K/s, and then switching to the NVE ensemble to
calculate dynamical properties.

For the calculation of the participation ratio and the function
$A_{\rm max}(r)$, equilibrium
configurations at a given temperature were quenched to obtain the
configurations corresponding to the nearest local minimum of the
system. Then we calculate eigenmodes and eigenvalues corresponding to
vibrational modes about this local energy minimum by diagonalizing the
Hessian matrix with respect to the generalized coordinates. We use the
flexible version of the model and hence all degrees of freedom
to calculate the Hessian
(see Refs.~\cite{Sastry1994,Matharoo2009,Bell1970} for a more detailed
explanation of the formalism and method).
 
\bibliographystyle{naturemag} 

 \section*{Acknowledgments} We thank S. V. Buldyrev and S. Sastry for helpful
   discussions. The simulations were in part performed using resources
   provided by the Swedish National Infrastructure for Computing (SNIC)
   at the NSC and HPC2N centers. LGMP, KTW and DS were supported by the
   Swedish Research Council. KTW is also supported by the Icelandic Research Fund through the START programme. PK acknowledges the support of National Academies Keck Future Initiatives award. HES thanks NSF Grants No. CHE0911389, No. CHE0908218, and No. CHE-1213217.

\begin{figure*}[htb]
\begin{center}
\includegraphics[width=10cm]{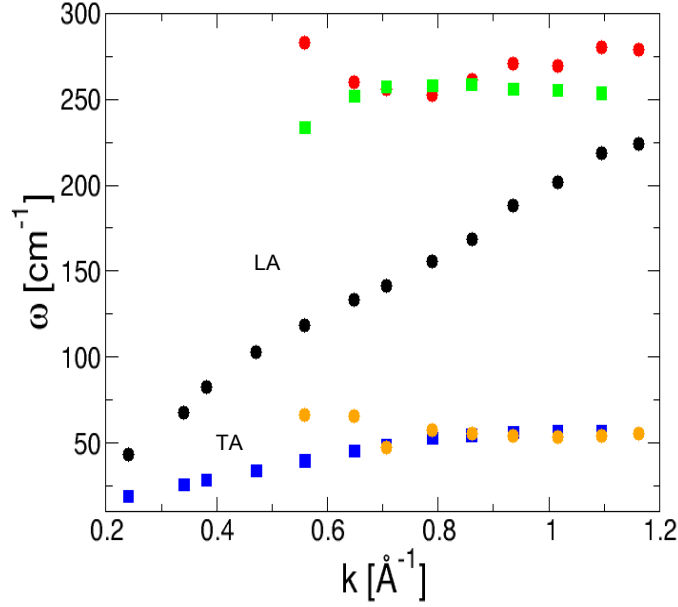}
\end{center} 
\caption{Longitudinal (filled circles) and transverse
  (filled squares) dispersion relations calculated from the peak
  positions of $\omega_{L}(k)$ in $C_L(k,\omega)$ and $\omega_{T}(k)$
  in $C_T(k,\omega)$ for N=$2048$ and T=210K. For $k<0.5$~\AA$^{-1}$, longitudinal spectra exhibit one dispersive branch (LA: black filled circles), while for $k>0.5$~\AA$^{-1}$,
  the longitudinal spectra exhibit three excitations --low and high
  frequency non-dispersive excitations (shown in orange and red
  filled circles). For $k<0.5$~\AA$^{-1}$ transverse current spectra exhibit only one dispersive excitation (TA: blue filled squares) while for $k>0.5$~\AA$^{-1}$, it exhibits both the dispersive branch as well as a non-dispersive excitation (greeen filled squares) at $\omega \approx 260$~cm$^{-1}$. For information on extraction of the dispersion relation see Fig.~S2. }
\label{fig:fig4} 
\end{figure*} 

\begin{figure*} 
\begin{center} 
\includegraphics[width=14cm]{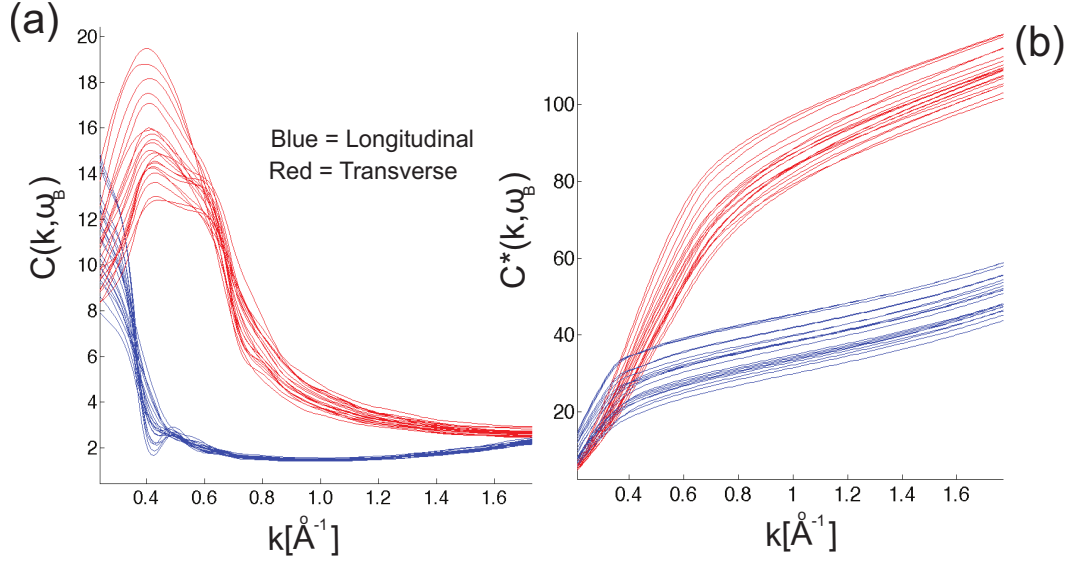} 
\end{center} 
\caption{(a) Current correlation functions
  $C_L(k,\omega_B)$ and $C_T(k,\omega_B)$ as function of $k$. (b)
  Integrated current correlation functions $C^*_L(k,\omega_B)$ and
  $C^*_T(k,\omega_B)$ as function of $k$ (see
  Eq.~\ref{eq:intcurr}). Several frequencies $\omega_B$ in the Boson
  peak frequency regime are shown. }

\label{fig:fig5}
\end{figure*}

\begin{figure*} 
\begin{center} 
\includegraphics[width=16cm]{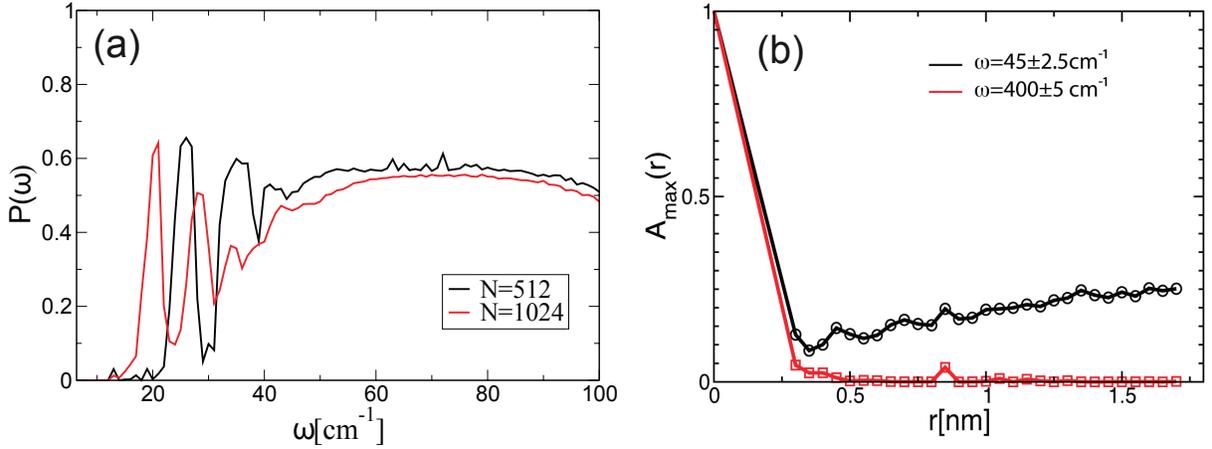} 
\end{center} 
\caption{(a) Frequency dependent participation ratio for two 
  system sizes, $N=512$ and $N=1024$. (b) Spatial dependence of the amplitude of normal
  modes $A_{\rm max}(r)$ in the Boson peak regime. For a comparison, 
  $A_{\rm max}(r)$ for $\omega=400 \pm 5cm^{-1}$ where the
  modes are localized is also shown. The value of $A_{\rm max}(0)$ is normalized to
  $1$ for the range of frequencies shown.}
\label{fig:fig7}
\end{figure*}

\clearpage

\section*{SUPPLEMENTARY INFORMATION~~~}

In Fig.~\ref{fig:figures1} we show the self-intermediate scattering function
$F_s(k,t)$ and the incoherent dynamic structure factor $S_s(k,\omega)$ of the
TIP5P water model which differs significantly from the TIP4P/2005 model in 
terms of physical properties. The Widom line in this model is close to 250~K,
which is also where the Boson peak in the model emerges.

In Fig.~\ref{fig:figures2} we show the transverse and longitudinal current
spectra $C_T(k,\omega)$ and $C_L(k,\omega)$ for TIP4P/2005 at
T=$210$~K for small, intermediate, and large wave numbers.

In Fig.~\ref{fig:figures3} we show a snapshot
of an energy-minimized configuration taken from a simulation at
$T=210$~K.  Molecules contributing the most to a given normal mode near
the peak frequency of the Boson peak are highlighted.
Specifically, we
looked at the molecules whose displacement is larger than a certain
percentage of the largest molecular displacement observed for the
given normal mode. We find no significant localization indicating that
the Boson peak corresponds to rather long-range collective motion in
the system.

Figure~\ref{fig:figures4} shows movies of normal modes associated with the
first two sharp peaks in (a) and (b) respectively, and the Boson peak in
(c).

\bigskip

\newpage

\setcounter{figure}{0}
\makeatletter
\renewcommand{\thefigure}{S\@arabic\c@figure}

\setcounter{figure}{0}
\makeatletter
\renewcommand{\thefigure}{S\@arabic\c@figure}

\begin{figure*} 
\begin{center} 
\includegraphics[width=16cm]{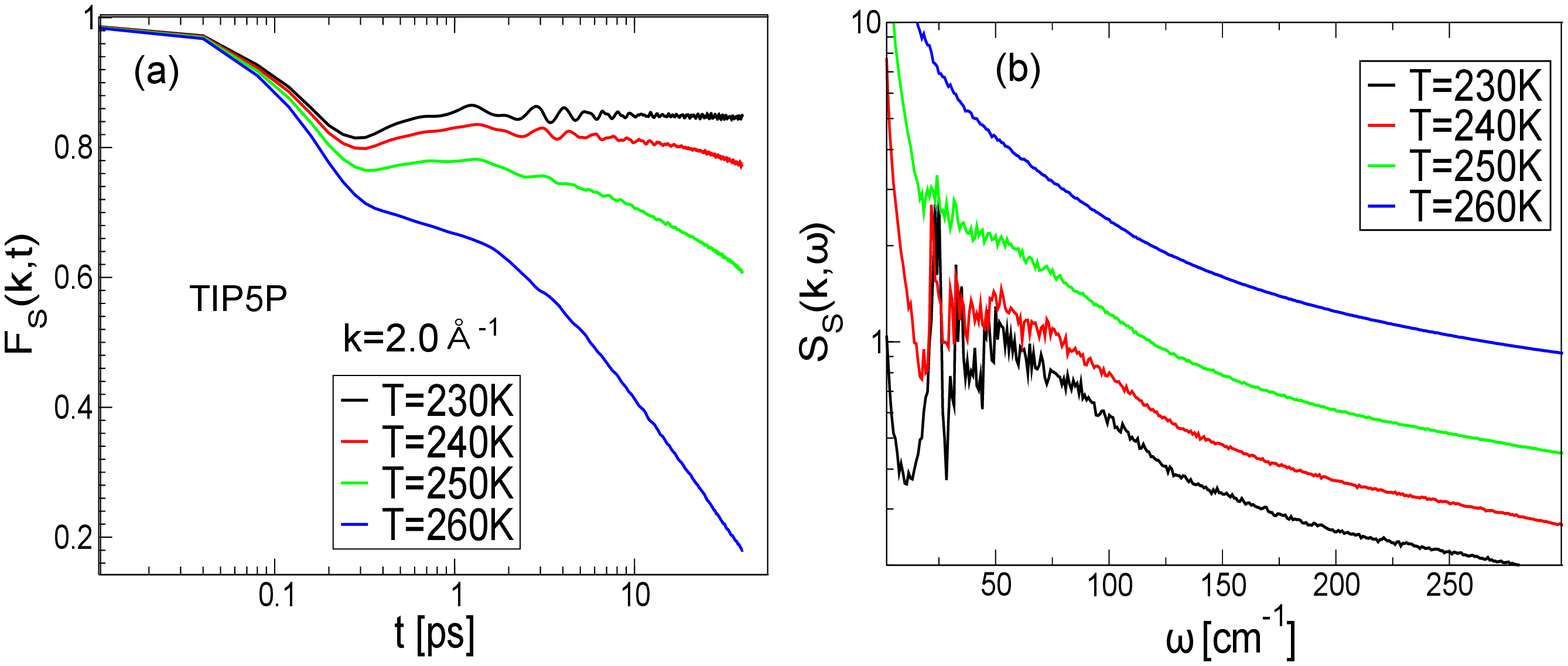} 
\end{center} 
\caption{(color online) (a) $F_s(k,t)$ and (b) $S_s(k,\omega)$ for 
  temperatures between 230 and 260~K at $k\approx 2.0$~{\AA}$^{-1}$ and system
  size $N=512$ for the TIP5P water model. Below the Widom line temperature of the model, 250~K, 
  a Boson peak around 50~cm$^{-1}$ is seen to emerge.}
\label{fig:figures1} 
\end{figure*} 
\clearpage

\begin{figure*} 
\begin{center} 
\includegraphics[width=12cm]{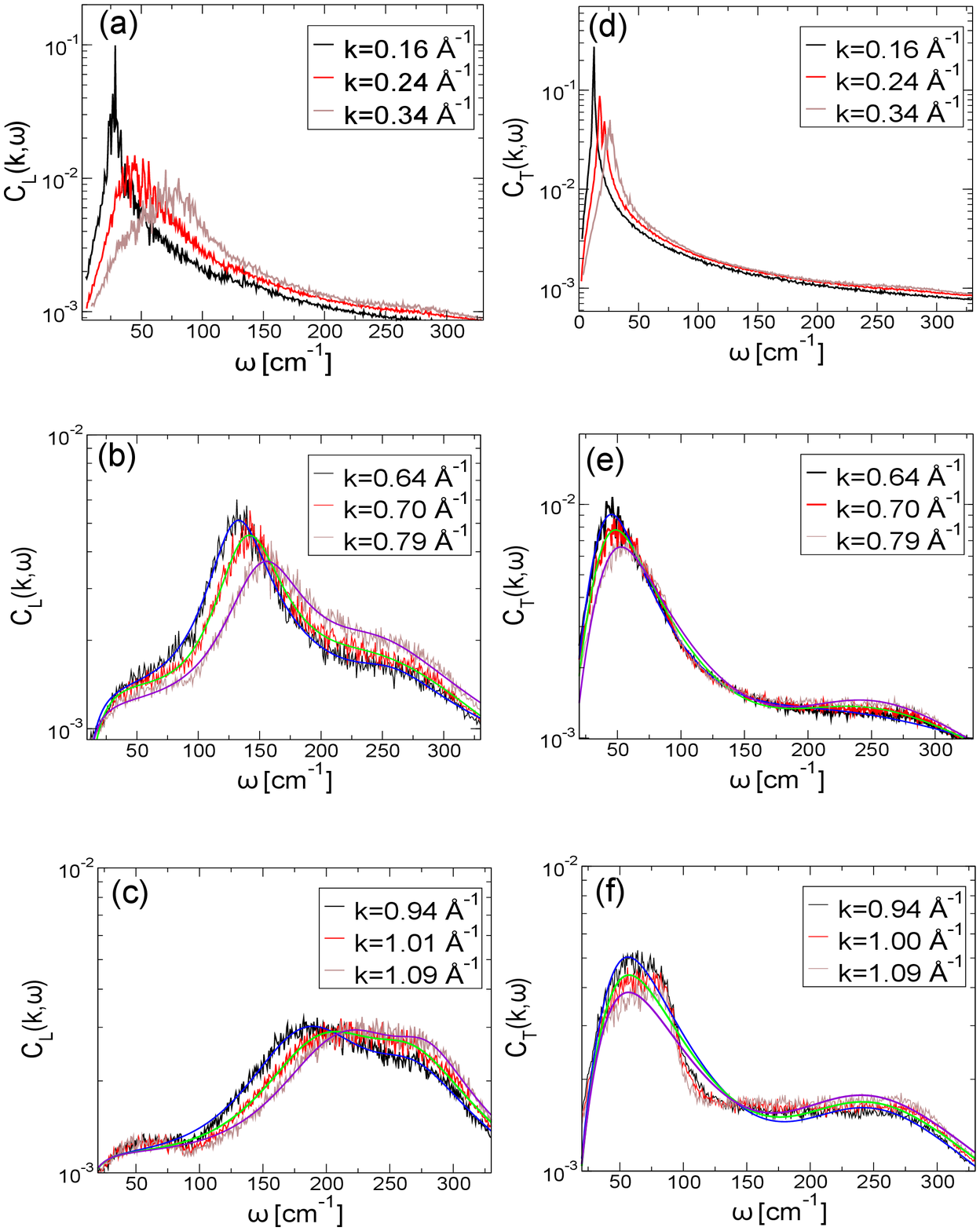} 
\end{center} 
\caption{(color online) Longitudinal current spectra $C_L(k,\omega)$
  at T=$210$~K for (a) small, (b) intermediate, and (c) large wave
  numbers. Besides the dispersing peaks $C_L(k,\omega)$ develops two non-dispersive excitations --one at low frequency $\omega \approx 50$~cm$^{-1}$ and a  high-frequency excitation at about $\omega=260$~cm$^{-1}$ for
  $k>0.5$~{\AA}$^{-1}$ and the propagating branch tends to saturate to
  this frequency at large wave numbers. Transverse current spectra
  $C_T(k,\omega)$ for (d) small (e) intermediate, and (f) large
  wave numbers at T=$210$~K. For $k<0.5$~\AA$^{-1}$, $C_T(k,\omega)$ exhibits one dispersing excitation but for $k>0.5$~\AA$^{-1}$ two excitations are observed. DHO fits to the spectra for $k>0.5$~\AA$^{-1}$ are shown in solid lines. For $k>0.5$~\AA$^{-1}$, longitudinal spectra are fit
  with three DHO excitations while the transverse spectra are fit with
  two DHO excitations. Note that for $k>0.5$~\AA$^{-1}$, longitudinal spectra exhibit both low and high frequency excitations corresponding to transverse excitations.}
\label{fig:figures2}
\end{figure*}
\clearpage

\begin{figure*}
\begin{center}
\includegraphics[width=16cm]{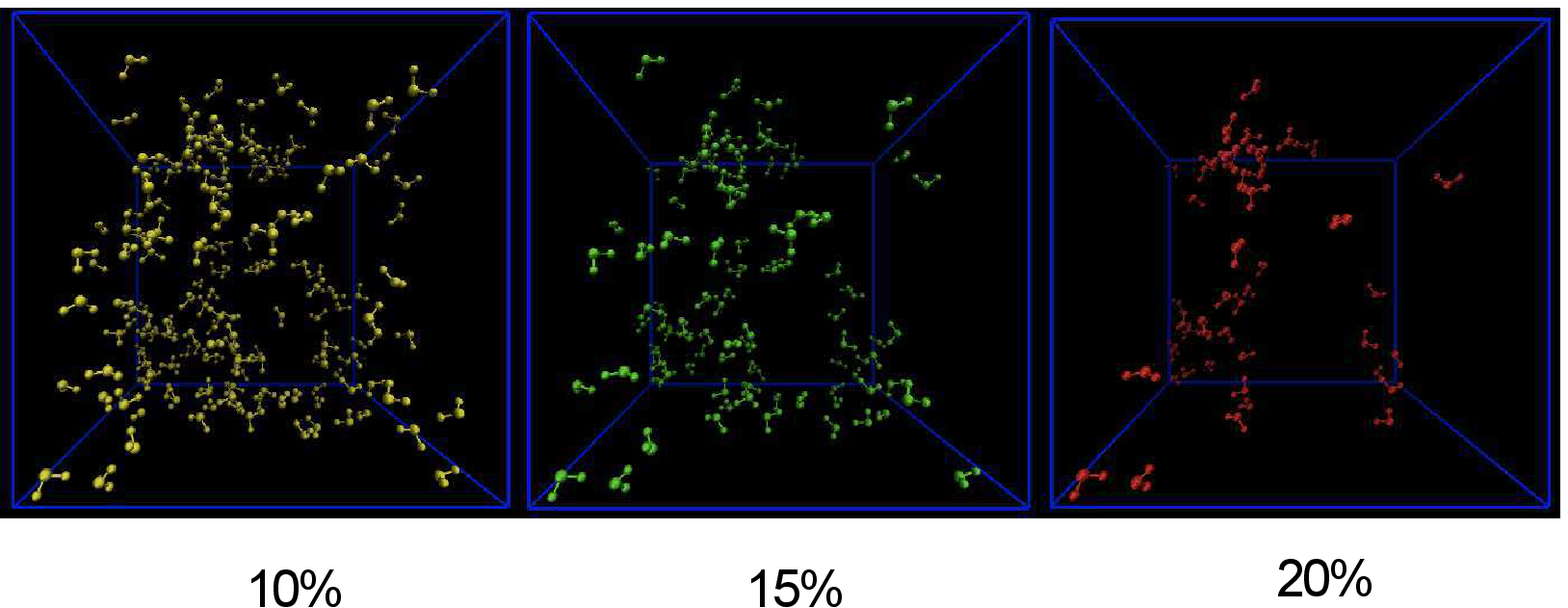}
\end{center} 
\caption{(color online) Snapshot of a configuration showing the molecules involved in a representative eigenmode at the Boson peak. (a) The molecules colored yellow are those with displacements larger than $10\%$ of the largest displacement of a molecule in the  normal mode. (b) The molecules colored green are those with displacement larger than $15\%$, and (c) molecules with displacement larger than $20\%$ are colored in red. We do not see significant localization in the low-frequency region where VDOS shows an excess.}
\label{fig:figures3} 
\end{figure*} 
\clearpage

\begin{figure*}
\begin{center}
\includegraphics[width=16cm]{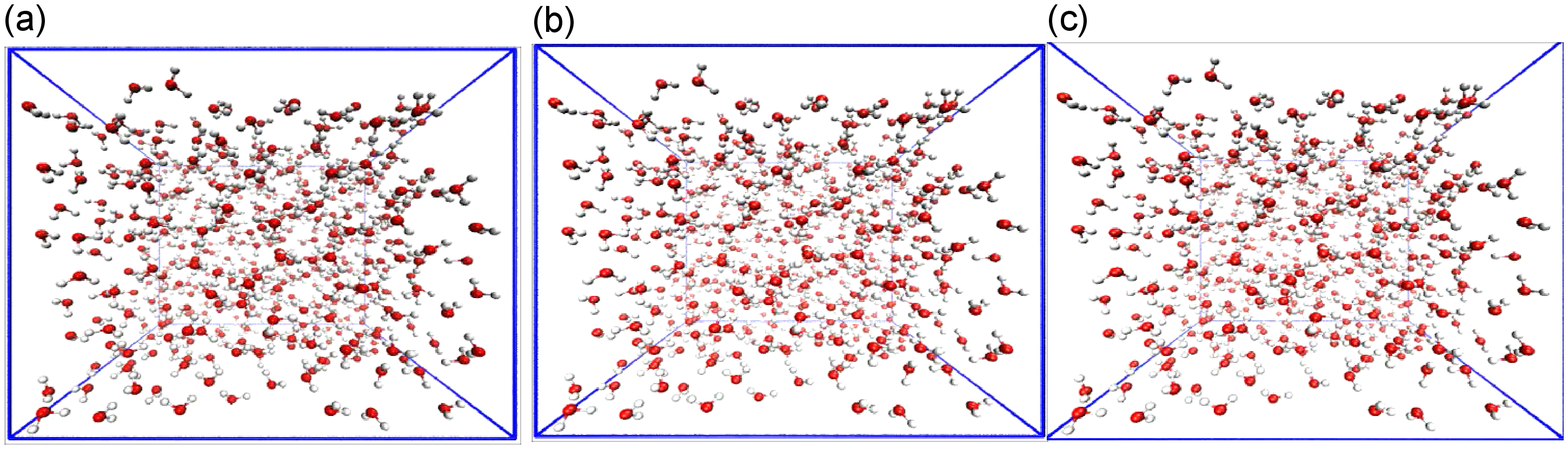}
\end{center} 
\caption{Movies of normal modes corresponding to the first two spurious size-dependent peaks (a) , (b) and (c) the normal mode corresponding to the position of the Boson peak. Please refer to the included video files nmtrajFreq1.mpg, nmtrajFreq2.mpg, and nmtrajFreqBP.mpg respectively.}
\label{fig:figures4}
\end{figure*}
\clearpage

\end{document}